\documentclass[conference]{IEEEtran}

\IEEEoverridecommandlockouts

\usepackage{cite}
\usepackage{amsmath,amssymb,amsfonts, amsthm}
\usepackage{algorithmic}
\usepackage{graphicx}
\usepackage{textcomp}
\usepackage[utf8]{inputenc}
\usepackage[T1]{fontenc}
\usepackage{multicol}
\usepackage{multirow}

\def\BibTeX{{\rm B\kern-.05em{\sc i\kern-.025em b}\kern-.08em
    T\kern-.1667em\lower.7ex\hbox{E}\kern-.125emX}}
    
\begin{document}

\title{Dynamic Structural Similarity on Graphs}

\author{\IEEEauthorblockN{1\textsuperscript{st} Eduar Castrillo}
\IEEEauthorblockA{\textit{Departamento de Ingeniería de Sistemas} \\
\textit{Universidad Nacional de Colombia}\\
Bogotá D.C., Colombia \\
emcastrillov@unal.edu.co}
\and
\IEEEauthorblockN{2\textsuperscript{st} Elizabeth León}
\IEEEauthorblockA{\textit{Departamento de Ingeniería de Sistemas} \\
\textit{Universidad Nacional de Colombia}\\
Bogotá D.C., Colombia \\
eleonguz@unal.edu.co}
\and
\IEEEauthorblockN{3\textsuperscript{st} Jonatan Gómez}
\IEEEauthorblockA{\textit{Departamento de Ingeniería de Sistemas} \\
\textit{Universidad Nacional de Colombia}\\
Bogotá D.C., Colombia \\
jgomezpe@unal.edu.co}
}

\maketitle

\begin{abstract}
One way of characterizing the topological and structural properties of vertices and edges in a graph is by using structural similarity measures. Measures like Cosine, Jaccard and Dice compute the similarities restricted to the immediate neighborhood of the vertices, bypassing important structural properties beyond the locality. Others measures, such as the generalized edge clustering coefficient, go beyond the locality but with high computational complexity, making them impractical in large-scale scenarios. In this paper we propose a novel similarity measure that determines the structural similarity by dynamically diffusing and capturing information beyond the locality. This new similarity is modeled as an iterated function that can be solved by fixed point iteration in super-linear time and memory complexity, so it is able to analyze large-scale graphs. In order to show the advantages of the proposed similarity in the community detection task, we replace the local structural similarity used in the SCAN algorithm with the proposed similarity measure, improving the quality of the detected community structure and also reducing the sensitivity to the parameter $\epsilon$ of the SCAN algorithm.

\end{abstract}


\section{Introduction}
Networks are ubiquitous because they conform the backbones of many complex systems, such like social networks, protein-protein interactions networks, the physical Internet, the World Wide Web, among others \cite{erciyes2014complex}. In fact, a complex system can be modeled by a complex network, in such a way that the complex network represents an abstract model of the structure and interactions of the elements in the complex system \cite{newman2011structure}. For example, in social networks like Facebook, the network of user-user friendships, the social groups and the user reactions to posts, can be modeled as complex networks.

It could be possible to predict the functionality or understand the behavior of a complex system if we get insights about it by analyzing its underlying network. For instance, if we are capables of detecting groups of vertices with similar topological features in the network, we can get insights about the particular roles played by each vertex (e.g. hubs, outliers) or how entire groups (e.g. clusters) describe or affect the overall behavior of the complex system \cite{erciyes2014complex}. The main problem that arise in this kind of analyses is how to determine efficiently high quality topological or structural similarities in the network. For this reason, several methods have been proposed to try to cope the problem. For example the local Degree centralitiy, the global Closeness centrality \cite{erciyes2014complex}, Betweeness centrality\cite{brandes2008variants} and Bridgeness centrality \cite{jensen2015detecting} characterize the importance of a vertex or an edge in a network. Also, there are local structural similarity measures between vertices like the Jaccard, Cosine or Dice that are based on the connectivity patterns of the vertices in their immediate neighborhood \cite{chang2017mathsf}. Additionally, there are more sofisticated methods like PageRank that ranks vertices in a network by using markov chain models \cite{pagerank}, or SimRank that computes the similarity between two vertices on directed graphs by using recursive similarity definitions \cite{jeh2002simrank}.

The structural similarity measures mentioned above, and other similars have been effectively used in graph clustering tasks \cite{castrillo2017fast, shao2014community, chang2017mathsf, xu2007scan, chen2012dense}. However, those similarities present a main drawback, i.e., those are limited to the immediate neighborhood of the connected vertices being measured. This limitation bypass important structural properties that can improve the quality of the structural similarity and therefore its applications. There are generalizations to compute structural similarities beyond the locality, like the one proposed in \cite{radicchi2004defining}, but their computation require algorithms with high computational complexity, making them impractical in large-scale scenarios.

The objective of this research work is to deal with the aforementioned drawback of the classical structural similarities. So, this paper presents a novel measure to compute the structural similarity of neighboring vertices based on the following intuitive definition: \emph{Two connected vertices are structurally similar if they share an structurally similar neighborhood}. Opposed to the classical structural similarities, our approach dynamically diffuses and captures information beyond the locality, describing the structural similarity of connected vertices in an entire graph at different points of time. This new similarity is modeled as an iterated function that can be solved by fixed point iteration in super-linear time and memory complexity, so it is able to analyze large-scale graphs. This new similarity exhibit interesting properties: \emph{i)} By definition, it is self-contained and parameter free; \emph{ii)} A dynamic interaction system emerges from it, because it exhibits many fluctuations for the similarities at early stages and converges to non-trivial steady-state as the system evolves. In order to show the advantages of the proposed similarity in the community detection task, we replace the local structural similarity used in the state-of-the-art SCAN algorithm with the proposed similarity measure, improving the quality of the detected community structure and also the sensitivity to the parameter $\epsilon$ of the SCAN algorithm.

The rest of this paper is organized as follows: In section \ref{sec_related_work} we briefly survey the related work about similarity measures. Section \ref{sec_dss} presents in detail the proposed dynamic structural similarity. Section \ref{sec_iscan} describes an application of the dynamic structural similarity in the task of community detection. In Section \ref{sec_experiments} several experiments are carried with the dynamic structural similarity. Finally, Section \ref{sec_conclusions} draws some conclusions about this research work.

\section{Related Work}
\label{sec_related_work}

Given a graph $G = (V, E)$, with set of vertices $V$, set of edges $(u, v) \in E$, a centrality, similarity or ranking measure is usually defined as a function $f:V\mapsto\mathbb{R}$ or $f:E\mapsto\mathbb{R}$, such that vertices and edges respectively are mapped to real values to determine their importance\footnote{the definition of importance depends on the specific problem being studied.} in the structure or functioning of the graph. A function $f$ can be classified into two main categories: local and global. This classification depends on how much information is required from $G$ in the computation of the function. Usually, in global functions the prior knowledge of the entire graph is required, otherwise the functions are considered local. Several centrality, similarity and ranking functions have been proposed in the literature, next we mention the state-of-the-art closely related to this research work.

\subsection*{Centrality measures}

The degree centrality is based on the idea that vertices with high degree are involved more frequently in communications than vertices with low degree. For a vertex $u$, the degree centrality is basically its degree. This is a local function, because it only depends on the neighborhood of $u$. Several definitions based on the degree centrality have been proposed, like k-path centrality and edge-disjoint k-path centrality. To compute these centrality measures several randomnized algorithms are proposed \cite{erciyes2014complex}.

The betweeness centrality is frequently used to measure the importance of vertices and edges in a graph. This centrality is based on the idea that the importance of a vertex/edge is proportional to the number of shortest paths passing by the investigated vertex/edge. The higher the betweenes centrality, the more important are vertices/edges for communication pourposes \cite{erciyes2014complex}. By definition this is a global function. So, the main limitation of the betweeness centrality is its computational cost, since it requires $O(|E|)$ computations per vertex/edge and $O(|V||E|)$ for the entire graph, making it impractical for very large graphs\cite{brandes2008variants, jensen2015detecting}. For this reason, several algorithms have been proposed to compute efficiently approximate values. Those algorithms are usually based on sampling methods or $(\delta, \epsilon)$ approximations \cite{bader2007approximating, riondato2016abra}.

\subsection*{Dynamic measures}

The core idea behind dynamic measures is the concept of random walk. A random walk is an iterative process that starts from a random vertex, and at each step, either follows a random outgoing edge of the current vertex or jumps to a random vertex. An algorithm based on the random walk intuition is PageRank (PR). PR \cite{pagerank} is a vertex ranking method that defines the importance of a vertex recursively as follows: \emph{The importance of a vertex in the network is proportional to the importance of the vertices pointing to it}. This algorithm models a random surfer who is placed in an specific web page and then navigates the web by clicking on links. However, the surfer starts to navigate from a random web page with a probability given by a damping factor $\alpha$ (tuned by hand). The PageRank is modeled as the stationary distribution of a markov chain process solved by fixed point iteration. Several dynamical systems have been proposed since the original PR, like Personalized PageRank (PPR) \cite{lofgren2016personalized}, Heat Kernels (HK) \cite{kloster2014heat} and pure random walks (RW) \cite{pons2005computing}. All of them compute similarities for seed vertices respect to the whole graph, hindering the simultaneously computation of multiple similarities.

Another popular dynamic similarity based on the random walk intuition is SimRank \cite{jeh2002simrank}. SimRank defines structural-context similarity of vertices (directly connected by edges or not) recursively as follows: \emph{Two objects are similar if they are related to similar objects}. The SimRank is modeled as a recursive function solved by fixed point iteration. By definition, SimRank only works for directed graphs, and also requires a decay factor $C$ in order to control the flow of information in the dynamic system and to achieve convergence.

An algorithm to compute structural similarity based on distance dynamics have been proposed in \cite{shao2014community}. They propose a fast algorithm to detect high-quality community structure by merging nodes with high structural similarity. The structural similarity is defined as the result of three interaction patterns that dynamically change the similarity through the time. These interaction patterns are solved by fixed point iteration. Although the system presents dynamic behavior, they force its convergence by truncating similarities above one, and below zero.

\subsection{Graph Model}

\theoremstyle{definition}
\newtheorem{defn}{DEFINITION}

\begin{defn}
Let $G = (V, E, \omega)$ be a graph with set of vertices $V$, set of edges $(u,v) \in E$ such that $u, v \in V$, and edge weighting function $\omega : E \mapsto \mathbb{R}$. In the case of undirected graphs, the edges $(u, v)$ and $(v, u)$ are considered the same. In the case of unweighted graphs $\omega(u, v) = 1$ for each $(u, v) \in E$. For the rest of this paper we suppose $G$ is an undirected and unweighted graph, unless other type of graph is explicitly mentioned.
\end{defn}

\begin{defn}
The structural neighborhood of a vertex $u$, denoted by $N(u)$, is defined as the open neighborhood of $u$; that is $N(u) = \{v \in V | (u, v) \in E\}$. Additionally, the closed structural neighborhood, denoted by $N[u]$, is defined as $N[u] = N(u) \cup \{u\}$.
\end{defn}

\begin{defn}
The degree of a vertex $u$, denoted by $d[u]$ and $d(u)$, is basically the cardinal of the structural neighborhood of $u$; that is $d[u] = |N[u]|$ and $d(u) = |N(u)|$.
\end{defn}

\subsection{Structural Similarity}

\begin{defn}
\label{def_structural_equivalence}
\textbf{\emph{Structural Equivalence}}: Two vertices in a network are structurally equivalent if they share the same neighborhood; that is,  given two vertices $u$ and $v$, then $N(u) = N(v)$. 
\end{defn}

However, computing $|N(u) \cap N(v)|$ is not considered a good similarity measure by itself, because it has not into account the degrees of the vertices. The structural equivalence can be improved by normalizing its value as the \emph{Structural Similarity} does.

\begin{defn}
\textbf{\emph{Structural Similarity (a.k.a Cosine Similarity)}}: The local structural similarity (LSS) of vertices $u$ and $v$, denoted by $\sigma(u, v)$, is defined as the cardinal of the set of common neighbors $|N[u] \cap N[v]|$, normalized by the geometric mean of their degrees, that is
\end{defn}

\begin{equation}
\label{ss_original}
	\sigma(u, v) = \frac{|N[u] \cap N[v]|}{\sqrt{d[u] \times d[v]}}
\end{equation}

By definition, the structural similarity is a local function because only requires information about the inmediate neighborhood of vertices $u$ and $v$. In fact, given two vertices $u$ and $v$, the structural similairy $\sigma(u, v)$ can be computed in $O(min(d[u], d[v]))$ time. Furthermore, $O(\alpha(G)\times|E|)$ time is required to compute the structural similarity for each pair of vertices in a graph $G$, such that the term $\alpha(G)$ corresponds to the arboricity of $G$ \cite{chiba1985arboricity}. The structural similarity is just an extension to the context of graphs of the Cosine Similarity. Another two popular similarity measures extended to the context of graphs are Jaccard and Dice, defined as $J(u, v) = |N[u] \cap N[v]| \div |N[u] \cup N[v]|$ and $D(u, v) = 2\times|N[u] \cap N[v]| \div (d[u] + d[v])$ respectively.

\section{Dynamic Structural Similarity}
\label{sec_dss}

The local structural similarity $\sigma(u, v)$ is a good quality measure, but it is limited to the immediate neighborhood of $u$ and $v$. This limitation bypass important structural properties given by patterns of connections beyond the locality (paths of length 2, 3, 4, ..., N between $u$ and $v$). One approach to solve the aforementioned limitation is by computing explicitly paths with length greater than one between the vertices $u$ and $v$. This computation is done by performing complete enumeration, like the local edge clustering coefficient proposed in \cite{radicchi2004defining}. This coefficient is defined for the edge $(u, v)$ as the number of cyclic structures of length $k$ that the edge $(u, v)$ belongs to, normalized by the total number of cyclic structures of length $k$ that can be build given the degrees of the vertices $u$ and $v$. Other similarity measures based on complete enumeration are the global closeness and betweeness centralities \cite{erciyes2014complex}, that count the number of all-pairs shortest paths running through the edge $(u, v)$. The disadvantage of these approaches is the high computational complexity required to enumerate the paths, making them impractical in large-scale graphs.

\subsection{Dynamic Structural Similarity Definition}

In order to compute structural similarity without doing complete enumeration, we propose a diffusion system to spread and capture structural similarity. Our approach is based on the following  intuition: \emph{Two vertices are structurally similar if they share an structurally similar neighborhood}. Let $DSS(u, v)$ be the \emph{Dynamic Structural Similarity} (DSS) of $u$ and $v$. Following the intuitive definition a recursive function $DSS(u, v)$ is as follows,

\begin{equation}
\label{dss_function}
	DSS(u, v) = \frac{\displaystyle\sum_{\substack{x \in N[u] \cap N[v]}}DSS(u, x) + DSS(v, x)}{\sqrt{\displaystyle\sum_{\substack{x \in N(u)}}DSS(u, x) \times \displaystyle\sum_{\substack{y \in N(v)}}DSS(v, y)}}
\end{equation}

Following the idea of the structural equivalence \ref{def_structural_equivalence}, we extend the similarity of two vertices $u$ and $v$, not only to the cardinal of their common neighborhood, but to their common similarity, i.e., the sum of the similarities of the edges connecting $u$ and $v$ to each of their common neighbors $x$. The higher the total similarity in the common neighborhood, the more structurally similar must be the vertices $u$ and $v$. The common similarity correspond to the numerator in the Equation \ref{dss_function}. Additionally, the common similarity is divided by the geometric mean of the total similarity in the neighborhood of $u$ and the total similarity in the neighborhood of $v$. Such division is done in order to get the relative importance of the common similarity respect to the entire neighborhood (denominator), similarly to the normalization performed in the local structural similarity (Equation \ref{ss_original}). 

Under this similarity definition, if the edge $(u, u) \in E$, then the vertex $u$ will present a similarity of 2 respect to itself, i.e., $DSS(u, u) = 2$. Also, from Equation \ref{dss_function} is easy to see that $DSS$ is symmetric, i.e., $DSS(u, v) = DSS(v, u)$. Although the structural equivalence is defined for any pair of vertices, we set as base case $DSS(u, v) = 0$ if $(u, v) \notin E$ (See section \ref{sub_complexity_analysis} for the explanation of this constraint). 

Intuitively, the term in the numerator contribute positively to $DSS(u, v)$, this contribution is directly proportional to the total similarity in the common neighborhood of $u$ and $v$. On the other hand, the term in the denominator contributes negatively to $DSS(u, v)$, this contribution is inversely proportional to the geometric mean of the total similarity in the neighborhood of $u$ and the total similarity in the neighborhood of $v$. In fact, the main negative contribution is given by the vertices that are not common neighbors of $u$ and $v$.

\subsection{Computing the Dynamic Structural Similarity}

DSS in a graph $G$ can be computed by fixed point iteration. Let $DSS_t$ be the iterated function defined from $DSS$ on iteration $t$. For each iteration $t$, $|V|^2$ entries $DSS_t(u, v)$ are maintained, $DSS_t(u, v)$ gives the similarity measure between vertices $u$ and $v$ on iteration $t$. The next iteration $DSS_{t+1}(u, v)$ is computed based on $DSS_t(u, v)$. In the initialization step (when $t = 0$), if the graph is unweighted, the value of each $DSS_0(u, v)$ must be assigned to an equal and positive similarity score $s$, and zero if the edge does not exists in $G$.

\begin{equation}
\label{iss_init}
 DSS_0(u, v) = 
  \begin{cases} 
   s & \text{if } (u, v) \in E \\
   0 & \text{if } (u, v) \notin E
  \end{cases}
\end{equation}

In the case of weighted graphs, the initial similarity can be set to the edge's weight, i.e., $DSS_0(u, v) = \omega(u, v)$.

To compute $DSS_{t+1}(u, v)$ from $DSS_t(u, v)$, the Equation \ref{dss_function} is adapted as follows,

\begin{equation}
\label{iss_function}
	DSS_{t+1}(u, v) = \frac{\displaystyle\sum_{\substack{x \in N[u] \cap N[v]}}DSS_t(u, x) + DSS_t(v, x)}{\sqrt{\displaystyle\sum_{\substack{x \in N(u)}}DSS_t(u, x) \times \displaystyle\sum_{\substack{y \in N(v)}}DSS_t(v, y)}}
\end{equation}

\subsection{Complexity Analysis}
\label{sub_complexity_analysis}
Let $E_s \subseteq |V|^2$ be the set of vertex pairs to whose DSS is being computed. Setting $DSS_0$ in the initialization step takes $O(|E_s|)$ time. Furthermore, if the adjacent list for each vertex is sorted, then $N[u] \cap N[v]$ can be computed in $O(min(d[u], d[v]))$ time. Thus, one iteration of the iterated structural similarity (Equation \ref{iss_function}) requires $\sum_{\substack{(u, v) \in E_s}} min(d[u], d[v])$ operations. In \cite{chiba1985arboricity} it has been proved that the number of operations performed per iteration in Equation \ref{iss_function} has an upper-bound of $2 \times \alpha(G) \times |E_s|$, such that $\alpha(G) \leq \sqrt{|E|}$ is the arboricity of $G$. Finally, $T$ iterations of Equation \ref{iss_function} are performed, resulting in a total complexity of $O(T\times\alpha(G)\times|E_s|)$ time. As opposed to other dynamic similarity measures \cite{Jin2011AxiomaticRO}, we set $DSS(u, v) = 0$ whenever $(u, v) \notin E$ in order to reduce the number of similarity computations from a maximum of $O(|V|^2)$ entries to $O(|E|)$ entries. This reduction becomes specially important on sparse graphs with $|E| = O(|V|)$. The memory complexity is $O(|V|+|E|)$.

\begin{figure}[t]
\includegraphics[width=90mm]{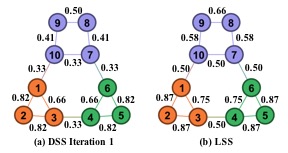}
\caption{First iteration of the dynamic structural similarity (a) and the local structural similarity (b) for all the edges in a toy network.}
\label{fig_dss_toy_net}
\end{figure}

\begin{table}[t]
\centering
\caption{Evolution of the Dynamic Structural Similarity in a toy network.}\label{table_dss_lss}

\begin{tabular}{ |c|c|c|c|c|c|c| }
\hline
Edge & \multicolumn{6}{ |c| }{Iteration} \\
\hline
     & 1 & 2 & 3 & 4 & 5 & $\geq$ 14 \\ \hline
 1-2 & 0.82 & 0.91 & 0.94 & 0.95 & 0.97 & 1.00 \\ \hline
 1-3 & 0.66 & 0.82 & 0.89 & 0.93 & 0.95 & 1.00 \\ \hline
 1-10 & 0.33 & 0.24 & 0.17 & 0.13 & 0.09 & 0.00 \\ \hline 
 2-3 & 0.82 & 0.91 & 0.95 & 0.97 & 0.99 & 1.00 \\ \hline
 3-4 & 0.33 & 0.18 & 0.10 & 0.05 & 0.02 & 0.00 \\ \hline
 4-5 & 0.82 & 0.91 & 0.95 & 0.97 & 0.99 & 1.00 \\ \hline
 4-6 & 0.66 & 0.82 & 0.89 & 0.93 & 0.95 & 1.00 \\ \hline
 5-6 & 0.82 & 0.91 & 0.94 & 0.95 & 0.97 & 1.00 \\ \hline
 6-7 & 0.33 & 0.24 & 0.17 & 0.13 & 0.09 & 0.00 \\ \hline
 7-10 & 0.33 & 0.31 & 0.32 & 0.35 & 0.38 & 0.50 \\ \hline
 7-8 & 0.41 & 0.41 & 0.43 & 0.45 & 0.46 & 0.50 \\ \hline
 8-9 & 0.50 & 0.55 & 0.57 & 0.57 & 0.56 & 0.50 \\ \hline
 9-10 & 0.41 & 0.41 & 0.43 & 0.45 & 0.46 & 0.50 \\ \hline
\end{tabular}
\centering
\end{table}

\section{ISCAN - Improved SCAN Algorithm}
\label{sec_iscan}

The structural similarity has been employed as a measure of cohesion within clusters and becomes specially useful to approximate dense subgraphs in networks with high \emph{Transitivity} and \emph{Community Structure} \cite{newman2011structure}, so it has been employed to perform community detection in complex networks \cite{castrillo2017fast, shao2014community, chang2017mathsf, xu2007scan, chen2012dense, han2016massivenets}. SCAN \cite{xu2007scan} is an algorithm that takes full advantage of the local structural similarity to perform community detection. SCAN is based on the idea of structure connected clusters expanded from seed core vertices. The core vertices are vertices with a minimum of $\mu$ adjacent vertices with structural similarity that exceeds a threshold $\epsilon$ \cite{xu2007scan}. An structure connected clusters $C_{\mu,\epsilon}$ is a maximal subset of vertices in which every vertex in $C$ is structure reachable from some core vertex in $C$. The SCAN algorithm requires two parameters: the minimum number of points $\mu$ and the minimum accepted similarity $\epsilon$, with default values of 2 and [0.5, 0.7] respectively according to the authors. Moreover, SCAN is one of the fastest algorithm in the literature, with total time complexity of $O(\alpha(G)\times|E|)$.

Because the SCAN algorithm performs quite well in terms of the quality of the resulting clustering, the majority of research works based on it, are focused to improve SCAN in terms of its computational complexity but not in the quality of the detected community structure or its usability. For example, the three recently proposed algorithms pSCAN \cite{chang2017mathsf}, SCAN++ \cite{shiokawa2015scan++} and Index-Based SCAN \cite{wen2017efficient} compute the same clustering results with optimized time complexities. In contrast, this research work is focused to improve the quality of the results and sensitivity to the parameterization of SCAN while maintaining the same computational complexity in practical cases. 

We consider that the key ingredient to improve the quality of the results and sensitivity of SCAN, is to support the cluster construction on a robust similarity measure. For that reason, the \emph{ISCAN} algorithm is proposed. ISCAN is obtained after replacing the local structural similarity used in SCAN with the proposed dynamic structural similarity. ISCAN presents a complexity of $O(T \times \alpha(G) \times |E|)$ time, given $T$ as the the number of iterations performed by the back-end DSS. Because $T$ does not scale with the size of network ($T \approx 5$, See \ref{subsub_lfr}), it can be considered a constant factor in most practical cases, therefore we argue that ISCAN keeps the same asymptotic time complexity of the original SCAN algorithm.

\section{Experiments}
\label{sec_experiments}
In this section we perform several experiments on real-world and synthetic benchmark graphs. The experiments are aimed to show the properties of the proposed similarity, and how it can be used to improve the overall performance of an algorithm that performs community detection on graphs.

\subsection*{Experimental Settings}
For all the experiments, the DSS is initialized with Equation \ref{iss_init} using an initial similarity score $s = 1$. In order to facilitate the comparison with the LSS, we normalize $DSS(u, v)$ to the interval $[0, 1]$ by applying min-max normalization, after performing the last required iteration of the Equation \ref{iss_function}.

In section \ref{sub_dynamic}, two real-wold networks extracted from \cite{leskovec2016snap} and a toy network are used to analyze the dynamic properties and evolution of the DSS through the time. In section \ref{sub_exp_iscan}, a set of synthetic benchmark graphs with planted ground-truth community structure is used to validate, evaluate and compare the proposed ISCAN algorithm.
 
\subsection{Evolution of the Dynamic Structural Similarity}
\label{sub_dynamic}

Fig. \ref{fig_dss_toy_net} shows the structural similarity computed in a toy network with both, the DSS and the LSS. As we can see, with one iteration of the DSS (Fig. \ref{fig_dss_toy_net}a), there are not important differences between both similarities (\ref{fig_dss_toy_net}b). However, the DSS deviates significantly from the LSS as the number of iterations increases, as shown in Table \ref{table_dss_lss}. In fact, the more number of iterations, the better characterization (in terms of the similarities) is obtained for both intra-cluster edges (e.g. 1-2, 4-6, 8-9, etc.) and inter-cluster edges (e.g. 1-10, 3-4, 6-7). Also, later iterations to iteration 14 not affect any similarity in the network, that means at iteration 14 the DSS achieves a non-trivial steady-state\footnote{A trivial steady-state is whose vertices are completely similar (maximal similarity for each edge) or completely dissimilar (minimal similarity for each edge).} in this sample network. 

\begin{figure}[t]
\includegraphics[width=90mm]{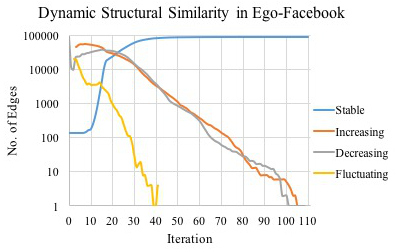}
\caption{Number of stable/increasing/decreasing/fluctuating edges (in log-scale) in the ego-facebook network \cite{leskovec2016snap} in function of the iteration of the Equation \ref{iss_function}. This plot shows the dynamic behavior of the proposed structural similarity in a real-world network. The majority of fluctuations occurs at early iterations and the system converge to a non-trivial steady-state.}
\label{fig_dss_ego_facebook}
\end{figure}

Because the toy network may not be informative enough about the dynamic properties, the DSS was applied in a real-world network. Fig. \ref{fig_dss_ego_facebook} shows the dynamic behavior of the DSS in an ego-network. The ego-network consist of friends lists from Facebook. This network was collected from survey participants using a Facebook app and is conformed by 4039 vertices and 88234 edges \cite{leskovec2016snap}. As we can see, the proposed structural similarity behave as a dynamical system in the ego-network, presenting many fluctuations (i.e., edges whose similarity decrease/increase in the iteration $t$ and increase/decrease on iteration $t+1$) at early iterations and reaching a non-trivial steady-state on later iterations\footnote{For the experiment with the ego-network, we consider stable any variation in the similarity below 1e-12. This threshold is no required to achieve the convergence, but it allows us to summarize the dynamic process with fewer iterations.}. This behavior differs from other dynamical similarities in the literature, consider for example the SimRank \cite{jeh2002simrank}, where the similarities increase monotonically after each iteration (no fluctuations occurs and no similarity decreases). We consider important a dynamical behavior because if the tendency in the similarity changes is known in advance, then the iterative process becomes less informative. Another interesting result from this experiment is the capability of our proposal to converge to a non-trivial steady-state without requiring a cooling factor (e.g. the constant $C$ in SimRank or the damping factor $\alpha$ in the Personalized Page Rank), this property allows the DSS to be parameter free.

In order to see with more detail the changes in the DSS through the time, a sample of fourteen edges was taken from the Youtube network \cite{leskovec2016snap}. The Youtube network is conformed by 1134890 vertices and 2987624 edges, each vertex represents an users and each edge represents an user-to-user friendship. 

\begin{figure}[t]
\includegraphics[width=90mm]{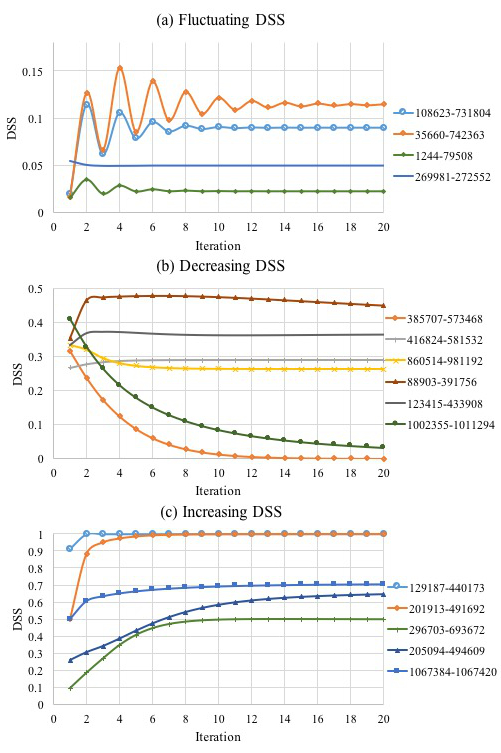}
\caption{Dynamic Structural Similarity in function of the iteration for a sample of fourteen edges taken from the Youtube network \cite{leskovec2016snap}. Each series corresponds to an edge in the network.}
\label{fig_changes_dss}
\end{figure}

Fig. \ref{fig_changes_dss} shows the changes in the DSS for each sample edge as the number of iterations increases. Several conclusions about the behavior of the DSS can be drawn: First, the initial values of the DSS can be loosely related to that values in later iterations, i.e., edges with high initial DSS can abruptly decrease (e.g. Fig. \ref{fig_changes_dss}b) and edges with low initial DSS can suddenly increase (e.g. Fig. \ref{fig_changes_dss}c). Second, The edges can fluctuate their DSS as the system evolves (e.g. Fig. \ref{fig_changes_dss}a and edges 88903-391756, 4168240-581532 in Fig. \ref{fig_changes_dss}b), with important fluctuations being observed at early iterations ($T \leq 16$). Third, the changes in the DSS occur at different rates, with some edges incresing/decreasing faster than others. Fourth, different edges with equal initial DSS not necessarily behave in the same manner as the system evolves (e.g. edges 416824-581532, 860514-981122 in Fig. \ref{fig_changes_dss}b and edges 201913-491692, 1067384-1067420 in Fig. \ref{fig_changes_dss}c). Fifth, all the sample edges tend to converge to a non-trivial steady-state.

\subsection{Performance of the ISCAN Algorithm}
\label{sub_exp_iscan}

\subsubsection*{\textbf{Evaluation Measures}} To test the effectiveness of ISCAN in the case of networks with ground truth communities, the sqrt-Normalized Mutual Information (NMI) have been selected. The NMI compares two partitions generated from the same dataset by assigning a score within the range [0, 1], such that 0 indicates that the two partitions are independent from each other and 1 if they are equal. On the other hand, if ground truth information is not provided, the performance of ISCAN is measured with the Modularity \emph{Q}. The modularity score is equal to $Q = 0$ if the partitions are not better than random, and is equal to $Q = 1$ if the partitions present strong community structure. In complex networks the Modularity usually falls in the interval $[0.3, 0.7]$, higher values to $Q = 0.7$ are rare and are biased towards the resolution limit \cite{newman2004finding, fortunato2007resolution}.

\subsubsection*{\textbf{LFR Benchmark}} LFR \cite{lancichinetti2008benchmark} generates unweighted and undirected graphs with ground-truth communities. Also, it produces networks with vertex degree and community size that follow power-law distributions, making it more appropriate than the Girvan-Newman benchmark to model complex networks. By varying the mixing parameter $\mu$, LFR can generate networks with community structure more or less difficult to identify. For our experiments, 30 LFR networks for each combination of the following parameters have been generated. We use networks with sizes $\{1000, 5000\}$. The average vertex degree is set to $k = 10$, which is of the same order in sparse graphs that represent real-world complex networks. The maximum degree is fixed to $kmax = 50$, and the community sizes vary in both small range $DSS = (10, 50)$ and big range $B = (20, 100)$. The vertex degree and community size follow power-law distributions with exponents $\tau1 = 2$ and $\tau2 = 1$ respectively. The mixing parameter $\mu$ vary in the interval $[0.05, 0.75]$ with step of $0.05$.

For the following experiments have into account that the parameter minimum number of points is fixed to $\mu = 0.2$ for both SCAN and ISCAN. Additionally, the number of iterations for the DSS in ISCAN is fixed to $T = 5$.

\subsubsection{Sensitivity to Parameter $\epsilon$}

The main drawback in the usability of the SCAN algorithm is the high sensitivity of the resulting clustering to the variations in the parameter $\epsilon$. This sensitivity was inherited from its predecessor algorithm DBSCAN\footnote{The SCAN algorithm is an adaptation to the context of graphs of the DBSCAN algorithm that originally performs density based clustering on spatial data.}, developed by the same author. 

\begin{figure}[t]
\includegraphics[width=90mm]{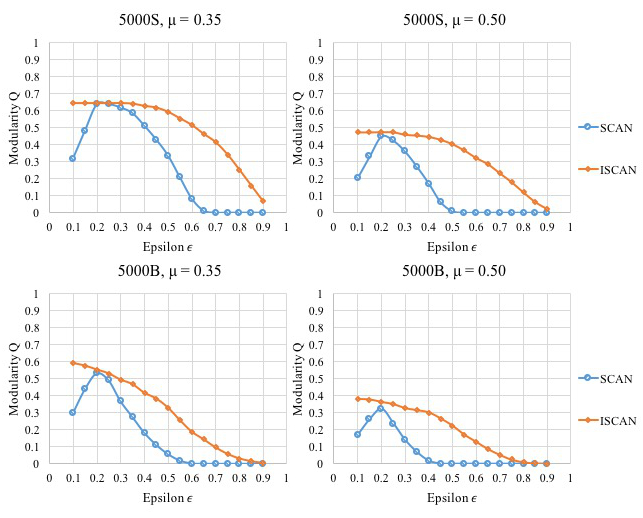}
\caption{Mean value of the Modularity score (higher values are better) in function of the parameter epsilon $\epsilon$ for the algorithms SCAN and ISCAN. The algorithms were executed on networks 5000S and 5000B with mixing parameters $\mu = \{0.35, 0.50\}$ (higher mixing parameter creates harder decision tasks).}
\label{fig_modularity}
\end{figure}

In order to compare the sensitivity to the parameter $\epsilon$ of SCAN and ISCAN, both algorithms have been executed over the 30 LFR networks generated with combination of parameters 5000S, 5000B and mixing parameters $\mu = \{0.35, 0.50\}$. For both, SCAN and ISCAN the parameter $\epsilon$ was varied in the interval $[0.1, 0.9]$ with steps of $0.05$. The mean and standard deviation of the Modularity score were computed to measure the quality of the results. As we can see in Fig. \ref{fig_modularity}, ISCAN presents better Modularity scores compared to SCAN for any variation of the parameter $\epsilon$. The SCAN algorithm increases effectively the Modularity from $\epsilon = 0.1$ up to $\epsilon = 0.2$, moment in which it achieves its maximum value, but for $\epsilon > 0.2$ SCAN shows an abrupt decay in its performance due to its high sensitivity to the parameter $\epsilon$. In contrast, ISCAN starts with maximum values of Modularity for $\epsilon = 0.1$ and decreases continuously as $\epsilon$ increases. However, ISCAN has mitigated the effects of the sensitivity to the parameter $\epsilon$.

\subsubsection{Detection of Ground-truth Communities}

In order to test the capacity of SCAN and ISCAN to detect ground-truth communities, both algorithms have been executed over the 30 LFR networks generated with each combination of parameters. The mean and standard deviation of the NMI were computed to measure the quality of the results. For both, SCAN and ISCAN the parameter $\epsilon = 0.2$ is fixed. The parameter $\epsilon$ was chosen based on the best average performance obtained in the benchmark graphs in Fig. \ref{fig_modularity}.

\begin{figure}[t]
\includegraphics[width=90mm]{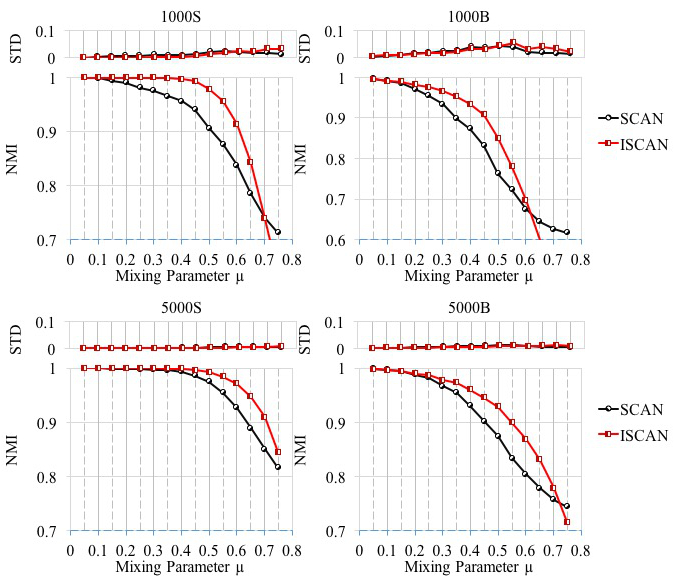}
\caption{(Lower row) Mean value of the sqrt-Normalized Mutual Information, NMI (higher values are better) as a function of the mixing parameter $\mu$. (Upper row) The standard deviation, STD (lower values are better) of the NMI as a function of $\mu$.}
\label{fig_lfr}
\end{figure}

As we can see in Fig. \ref{fig_lfr}, the critical point on the performance for both algorithms arrives when the mixing parameter $\mu \geq 0.5$. Anyway, ISCAN surpass the quality of the results of SCAN in the majority of scenarios, with up-to 8\% of increase in the quality of the results (NMI). Also, both algorithms remain stable for fixed parameters, presenting very low standard deviation, with $STD \leq 0.05$ in all scenarios. 

In order to test the sensitivity to the parameter $\epsilon$ of SCAN and ISCAN in the detection of ground-truth communities, both algorithms have been executed over the 30 LFR networks generated with combination of parameters 1000S, 1000B. For both, SCAN and ISCAN the parameter $\epsilon$ was varied in the interval $[0.2, 0.5]$ with steps of $0.1$. The mean of the NMI score was computed. 

\begin{figure}[t]
\includegraphics[width=90mm]{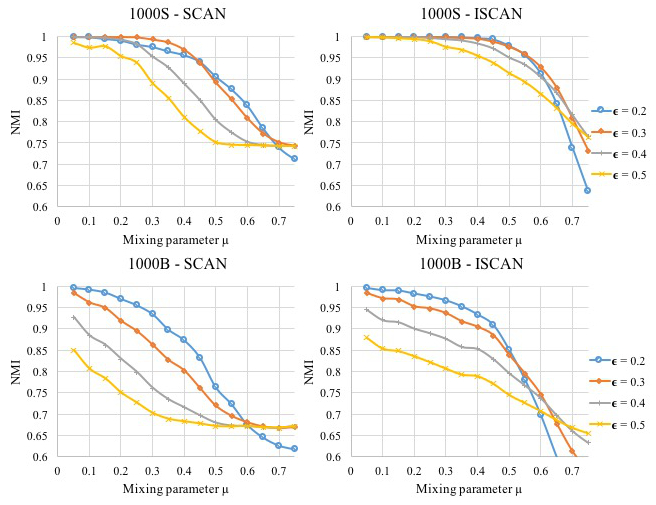}
\caption{Mean value of the sqrt-Normalized Mutual Information, NMI (higher values are better) as a function of the mixing parameter $\mu$. Series are plotted for different values of the parameter $\epsilon$ in the interval $[0.2, 0.5]$.}
\label{fig_eps_vs_nmi}
\end{figure}

As we can see in Fig. \ref{fig_eps_vs_nmi}, the quality of the results obtained with ISCAN are less sensitive compared to SCAN for any variation of the parameter $\epsilon$. Even though the NMI drops for both algorithms as $\epsilon$ increases, the NMI drops more rapidly in the case of SCAN.
 
\subsubsection{Community Size Distribution}

A recurrent problem in the community structure generated by the SCAN algorithm is the high number of singleton communities (communities with size 1) that are detected as hubs or outliers even if they do not exhibit such characteristic. In order to compare the community size distributions generated by SCAN and ISCAN, both algorithms have been executed on a sample\footnote{We take a sample network, but the experimental results follow the same trend in the remaining 29 networks.} network 5000S with mixing parameter $\mu = 0.5$ taken from the 30 networks generated with the LFR benchmark. In this sample network the community size distribution is within the range $[10, 50]$. For both, SCAN and ISCAN the parameter $\epsilon$ was tested in $\{0.2, 0.35\}$. The Mean Square Error (MSE) of the estimated community size distribution respect to the real distribution was computed.

\begin{figure}[t]
\includegraphics[width=90mm]{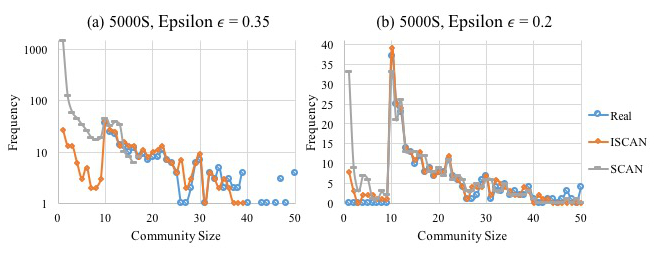}
\caption{The real community size distribution and the estimations given by the algorithms on a sample network 5000S with mixing parameter $\mu = 0.5$. The real community size distribution contains community sizes within the range [10, 50].}
\label{fig_com_size_dist}
\end{figure}

Fig. \ref{fig_com_size_dist}a shows the community size distribution generated by SCAN and ISCAN with parameter $\epsilon = 0.35$. For this parameterization ISCAN obtains NMI score of $0.99$ and SCAN obtains NMI score of $0.87$. SCAN generates a community size distribution within the range $[1, 17]$ with more than $1400$ singleton communities, resulting in a poor approximation compared to the real community size distributionwith MSE of $42194.56$. On the other hand, ISCAN generates community size distribution in the range $[1, 39]$ giving a better approximation to the real community size distribution with MSE of $26.14$. Fig. \ref{fig_com_size_dist}b shows the community size distribution generated by SCAN and ISCAN with parameter $\epsilon = 0.2$. In this case ISCAN and SCAN obtain NMI score of $0.99$ and $0.97$ respectively. Also, SCAN reduces the number of singleton communities to 33 but they are still high compared to the 8 singleton communities generated by ISCAN. In this configuration, SCAN obtains MSE of $28.66$ and ISCAN obtains MSE of $3.28$. This experiment evidence one more time the high sensitivity of SCAN to small variations in the parameter $\epsilon$, and how such sensitivity is mitigated with ISCAN. Moreover, in both cases ISCAN presents better estimation than SCAN of the ground-truth community size distribution.
 
\section{Conclusions}
\label{sec_conclusions}
In this paper a novel Dynamic Structural Similarity on graphs is proposed. It determines the structural similarity of connected vertices in a graph by using an iterated function that dynamically diffuses and captures structural similarity beyond the immediate neighborhood. Based on the experimental results, we claim that the Dynamic Structural Similarity is in fact a dynamical system, with many fluctuations in the similarities at early periods of time and convergence to non-trivial steady-state as the system evolves. Moreover, we show an application of the Dynamic Structural Similarity in the Community Detection task with the proposed ISCAN algorithm. Thanks to the back-end DSS, ISCAN outperforms the quality of the detected community structured and also reduces the sensitivity to the parameter $\epsilon$ compared to SCAN algorithm. ISCAN is achieved after replacing the local structural similarity used in the SCAN algorithm with our proposed dynamic structural similarity. As future work we plan to extend the Dynamic Structural Similarity to be computed on dynamic graphs.

\end{document}